\documentclass[aps,11pt,prd,groupedaddress,nofootinbib,notitlepage,eqsecnum,preprintnumbers]{revtex4-2}\usepackage[utf8]{inputenc}
\usepackage{hyperref}
\usepackage{xcolor}
\usepackage{graphicx}
\usepackage{amsmath,amssymb}
\usepackage{bm}
\usepackage{comment}
\usepackage[shortlabels]{enumitem}
\usepackage{overpic}
\usepackage{ulem}
\usepackage{orcidlink}
\usepackage{tensor}

\newcommand{\ba}{\begin{align}}
\newcommand{\ea}{\end{align}}

\begin{document}
\title{On the adiabatic initial conditions for a particle gas in cosmology}

\author{\textsc{Guillem Dom\`enech\,\orcidlink{0000-0003-2788-884X}}}
    \email{{guillem.domenech}@{itp.uni-hannover.de}}
\affiliation{Institute for Theoretical Physics, Leibniz University Hannover, Appelstraße 2, 30167 Hannover, Germany.}
\affiliation{ Max Planck Institute for Gravitational Physics, Albert Einstein Institute, 30167 Hannover, Germany.}

\begin{abstract}
In view of recent interest in the role of ``dark'' radiation  in cosmology, such as cosmic gravitational waves, sterile neutrinos and dark photons, we clarify the definition of adiabatic initial conditions in the kinetic theory of gases in an expanding universe. Without assuming any form for the phase space distribution function, we identify two possibilities: a \textit{strong} and a \textit{weak} adiabatic initial condition. The strong one corresponds to the standard adiabatic initial conditions, while the weak one is related to the strong via \textit{internal isocurvature} fluctuations. We show that both types of adiabatic initial conditions are consistent with the separate universe approach, although the latter requires initial internal isocurvature. In passing, we stress the importance of using the particle local momentum in the phase space to define the notion of adiabatic initial conditions. Doing so, we clarify that a gas of gravitons can have adiabatic initial conditions.
\end{abstract}
\maketitle

\section{Introduction\label{sec:intro}}

In the Hot Big Bang model, the Universe starts at high energy scales and in thermal equilibrium with tiny energy density fluctuations. Cosmic Microwave Background (CMB) observations measured such primordial fluctuations to be Gaussian, almost scale invariant, and adiabatic \cite{Planck:2018vyg}. Cosmic inflation \cite{Starobinsky:1979ty, Sato:1981qmu, Guth:1980zm, Linde:1981mu, Albrecht:1982wi} is the leading explanation for the initial conditions of the Hot Big Bang \cite{Planck:2018jri}. There, the nearly de Sitter expansion stretches quantum vacuum fluctuations to macroscopic scales \cite{Kodama:1984ziu,Mukhanov:1990me,Baumann:2009ds,Sasaki:2012ss}. These primordial fluctuations are the seeds of, e.g, CMB temperature fluctuations and  large scale structures we see in the Universe, like galaxies.

The natural expectation in single-field inflation, where only one scalar field drives inflation and later decays into standard model particles, is that primordial fluctuations are adiabatic. By adiabatic, it is meant that the initial conditions for different fluid fluctuations share a common origin and have no relative number density fluctuations. More precisely, there are no initial isocurvature fluctuations \cite{Kodama:1984ziu,Langlois:2003fq,Malik:2004tf,Malik:2008im}. At least, none is found on CMB scales \cite{Planck:2018jri}. 

Inflation also generates Gravitational Waves (GWs), see, e.g., Ref.~\cite{Guzzetti:2016mkm} for a review. Although no primordial GWs have so far been found in CMB scales \cite{Planck:2018jri}, higher frequency GWs effectively act as additional relativistic particles in the post-inflationary universe. In this way, one can also constrain the amount of cosmic GWs in searches of extra relativistic species in the CMB \cite{Planck:2018jri} and Big Bang Nucleosynthesis (BBN) analysis \cite{Cyburt:2004yc}. Note that, the constraints derived from CMB analysis vary depending on whether the initial conditions for GW density fluctuations are adiabatic or isocurvature \cite{Smith:2006nka,Clarke:2020bil}. Although one naively expects that GWs from single-field inflation are adiabatic, recent studies have questioned this \cite{ValbusaDallArmi:2023nqn,ValbusaDallArmi:2024hwm}. We will show that, within the graviton gas approximation, such naive expectation still holds.

GWs are particularly interesting. Extreme processes in the early universe, such as first-order phase transitions, cosmic strings and large primordial fluctuations, may have generated a detectable cosmic GW background \cite{Binetruy:2012ze,Caprini:2018mtu,Domenech:2021ztg,Roshan:2024qnv,Bian:2025ifp}. The cosmic GW background can be treated, in the high frequency limit, as a gas of gravitons \cite{Contaldi:2016koz,Bartolo:2019oiq,Bartolo:2019yeu}.\footnote{
Note that it is not completely clear when super-Hubble tensor fluctuations start to contribute to the energy density as GWs. At the moment, this seems to be couple of e-folds after Hubble radius entry  \cite{Smith:2006nka,Pritchard:2004qp}.} One then considers the Isaacson prescription for the GW energy momentum tensor \cite{Isaacson:1968hbi,ValbusaDallArmi:2024hwm} (see also Refs.~\cite{Giovannini:2019ioo,Negro:2024iwy}) to derive the energy density carried by the GWs and the corresponding graviton distribution function. In the graviton gas picture, one derives GW background anisotropies via the Boltzmann equation \cite{Contaldi:2016koz,Bartolo:2019oiq,Bartolo:2019yeu}. For applications, see Refs.~\cite{Ricciardone:2021kel,Bartolo:2019zvb,ValbusaDallArmi:2020ifo,Mierna:2024pkh}. However, there are currently two different prescriptions in the literature for adiabatic initial conditions for the graviton distribution function \cite{Ricciardone:2021kel,Dimastrogiovanni:2022eir,Malhotra:2022ply,Schulze:2023ich}. Here, we provide a complete classification of all the possible adiabatic initial conditions for a gas of relativistic particles.

The same discussion applies to more general situations. For instance, although there is no evidence for additional relativistic species \cite{Cyburt:2004yc,Planck:2018jri}, we have plenty of theoretical motivation to look for them. From the generation of neutrino masses \cite{Dolgov:2002wy,King:2015aea} and the matter-antimatter asymmetry \cite{Davidson:2008bu,Riotto:1999yt,Canetti:2012zc} to the nature of dark matter \cite{Bertone:2016nfn,Balazs:2024uyj}, all points to new physics beyond the standard model. Furthermore, dark radiation may alleviate the so-called Hubble tension \cite{Wyman:2013lza,Dvorkin:2014lea,Leistedt:2014sia,Graef:2018fzu,Archidiacono:2019wdp,Kreisch:2019yzn,Schoneberg:2019wmt,Blinov:2020hmc,Gariazzo:2023hch,Buen-Abad:2024tlb} (see also Ref.~\cite{DiValentino:2021izs,Vagnozzi:2023nrq,Hu:2023jqc} for recent reviews). And dark matter particles may have been relativistic at generation, as in ultra-light dark matter \cite{Ferreira:2020fam}, sterile neutrinos \cite{Balazs:2024uyj}, axions \cite{Chadha-Day:2021szb}, or dark photon dark matter \cite{Fabbrichesi:2020wbt,Cyncynates:2024yxm}.

Understanding the initial conditions is crucial for deriving the theoretical predictions. In this paper, we look in detail at the theoretical definition of adiabatic initial conditions regardless of the nature of the relativistic particles, i.e., whether they achieved a thermal spectrum or not, using the Kinetic Theory approach to a gas of particles in an expanding Universe, see Refs.~\cite{Andreasson:2011ng} and \cite{Choquet-Bruhat} for reviews. For applications to the CMB fluctuations, see Ref.~\cite{Ma:1995ey,Bruni:1996im} for an early works and Refs.~\cite{Pitrou:2007jy,Naruko:2013aaa,Saito:2014bxa,Namikawa:2021obu} for more recent ones. We find, by requiring the absence of isocurvature \cite{Malik:2004tf} and the separate universe approach \cite{Wands:2000dp}, that there are two general possibilities: a \textit{strong} adiabatic initial condition, which is frequency independent, and a \textit{weak} adiabatic initial condition, which is related to the strong adiabatic one via frequency dependent \textit{internal isocurvature} fluctuations.

This paper is organized as follows. In Sec.~\ref{sec:Kinetictheory}, we review cosmological perturbation theory in the Kinetic Theory of gases. There we closely follow Refs.~\cite{Pitrou:2007jy,Naruko:2013aaa,Saito:2014bxa}, which employ the local tetrad formalism. In passing, we emphasize that, to avoid misinterpretations, it is more appropriate to define adiabatic initial conditions using the local particle momentum. Then, in Sec.~\ref{sec:adiabatic_conditions} we investigate the definition of adiabatic initial conditions from the presence of a common uniform density slice (in Sec.~\ref{sec:common_uniform}), which is equivalent to the absence of isocurvature, and from the separate universe approach (in Sec.~\ref{sec:separate_universe}). We conclude our paper in Sec.~\ref{sec:conclusions} with some discussions. Throughout this work, we use natural units, that is $\hbar=c=1$, and the $(-,+,+,+)$ signature of the metric. We also use Greek letters for spacetime indices, Roman letters for spatial indices and capital Roman letters for local Lorentz indices.

\section{Review: Kinetic theory of gases in cosmology\label{sec:Kinetictheory}}

Our starting point is the phase space in the local inertial frame of the particles, which is built from the coordinates $x^\mu$ and the local conjugate momentum $P_A$, properly defined in the tangent space \cite{Pitrou:2007jy,Naruko:2013aaa,Choquet-Bruhat}. We define the local inertial frame through the \textit{tetrads} $\mathbf{e}_{(A)}$ in the tangent space (see Ref.~\cite{Yepez:2011bw} for a review) such that their inner product yields $(\mathbf{e}_{(A)},\mathbf{e}_{(B)})=\eta_{AB}$, where $\eta_{AB}$ is the Minkowksi metric. We denote the dual tetrads as $\mathbf{e}^{(A)}$. We then decompose the 4-momentum of the particle in the non-coordinate basis as $\mathbf{P}=P_A\mathbf{e}^{(A)}=P^A\mathbf{e}_{(A)}$.

The number of particles in an infinitessimal phase space volume reads \cite{Ma:1995ey,Choquet-Bruhat}
\begin{align}
dN=f(x^\mu,P_I)dx^1dx^2dx^3dP_{A_1}dP_{A_2}dP_{A_3}\,,
\end{align}
where $f(x^\mu,P_I)$ is the particle distribution function and we used $A_i$ with $i=\{1,2,3\}$ to denote the ``spatial'' components of the conjugate local momentum $P_A$, that is, the components of $P_A$ along the space-like tetrads. In the local inertial frame, the energy-momentum tensor is defined by \cite{Ma:1995ey}
\begin{align}\label{eq:TAB}
T_{AB}&=\frac{1}{2}\int dP_{A_0}dP_{A_1}dP_{A_2}dP_{A_3}\,f(x^\mu,P_I){P_AP_B}\,\delta^{(4)}(P_CP^C)\nonumber\\&=\int dP_{A_1}dP_{A_2}dP_{A_3}\,\frac{P_AP_B}{|P_{A_0}|}f(x^\mu,P_I)\,,
\end{align}
which is manifestly local Lorentz invariant. In Eq.~\eqref{eq:TAB}, $P_{A_0}$ is the ``time'' component of $P_A$, that is, the component along the time-like tetrad. Furthermore, since we focus on massless particles, we have that
\begin{align}
\eta^{AB}P_AP_B=-P_{A_0}^2+\delta^{{A_i}{A_j}}P_{A_i}P_{A_j}=0\,.
\end{align}
Thus, we may write for simplicity that
\begin{align}\label{eq:Phatai}
P_{A_i}=P^{A_0}\hat P_{A_i}\quad{\rm and}\quad P^{A_i}=P^{A_0}\hat P^{A_i}\,,
\end{align}
where normalized with respect to $P^{A_0}$ because it corresponds to the observed momentum of the particle, as we shortly show. Note that $P^{A_0}=-P_{A_0}$.
The distribution function $f$ can then be written as $f=f(x^\alpha,P^{A_0},\hat P^{A_i})$. This simplifies later calculations significantly.

We now promote the expressions in the local inertial frame to spacetime components. To do that, we write the tetrads in terms of the canonical spacetime basis given by $\mathbf{e}_{(\mu)}=\partial/\partial x^\mu$ and their dual $\mathbf{e}^{(\mu)}=d\mathbf{x}^{\mu}$. Namely, we write $\mathbf{e}_{(A)}=\tensor{e}{_A^\mu}(x^\alpha) \mathbf{e}_{(\mu)}$ and, conversely, $\mathbf{e}^{(A)}=\tensor{e}{^A_\mu}(x^\alpha) \mathbf{e}^{(\mu)}$. We denote the tensors $\tensor{e}{^A_\mu}$ as \textit{vierbeins} \cite{Yepez:2011bw}. The vierbeins are related to the spacetime metric via $g_{\mu\nu}=\tensor{e}{^A_\mu}\tensor{e}{^B_\nu}\eta_{AB}$ and similarly for their inverse. It then follows that the spacetime components of the particle momentum are $p_\mu=\tensor{e}{^A_\mu}P_A$ and $p^\mu=\tensor{e}{_A^\mu}P^A$. There is, however, some freedom to fix the form of the vierbeins, related to local Lorentz invariance. 

We  fix the vierbeins as in Refs.~\cite{Pitrou:2007jy,Naruko:2013aaa}. First, we choose a time-slice specified by a 4-vector $n^\mu$ orthogonal to a given spatial hypersurface, as done, e.g., in the 3+1 (ADM) decomposition. The proper time of a geodesic observer in that time-slice is given by $d\tau=(-n_\mu dx^\mu)$. Requiring the observed momentum to be equal to the local one, namely $(-n_\mu p^\mu)=P^{A_0}$, leads to $\tensor{e}{_{A_0}^\mu}=n^\mu$. From there, the orthogonality condition of the vierbeins, that is $n_\mu\tensor{e}{_{A_i}^\mu}=0$, is trivially satisfied if $\tensor{e}{_{A_i}^0}=0$. We fix the remaining freedom under Lorentz rotations after specifying the background vierbeins. We do so later, after writing down the explicit form of the spacetime metric. 

For later use, we also write the energy-momentum tensor in spacetime components, which reads
\begin{align}\label{eq:Tmunu}
T_{\mu\nu}=e^A_\mu e^B_\nu T_{AB}&=\frac{1}{2}\int \frac{dp_0dp_1dp_2dp_3}{\sqrt{-g}}f(x^\sigma,p_\beta){p_\mu p_\nu}\,\delta^{(4)}(p_\alpha p^\alpha)\nonumber\\&
=\int \frac{dp_1dp_2dp_3}{\sqrt{-g}}\frac{p_\mu p_\nu}{|p_0|}f(x^\sigma,p_\beta)\,,
\end{align}
where we used that $p_\mu=e^A_\mu P_A$ and $dp_0dp_1dp_2dp_3=(\det e) dP_{A_0}dP_{A_1}dP_{A_2}dP_{A_3}$. Note that, by definition, $\det e=\sqrt{-g}$. The distribution function $f$ is a scalar quantity and, therefore, only its arguments change, that is $f(x^\sigma,p_\beta)=f(x^\mu,P_I)$.

\subsection{Boltzmann equation}

Let us review the derivation of the Boltzmann equation in a general manner. The reader my find the lecture notes by Bertschinger \cite{Bertschinger:1993xt} and Sasaki \cite{SasakiCPTlectures} quite useful. For simplicity, we neglect particle interaction as its inclusion does not affect our later discussion on the adiabatic initial conditions. Now, since the phase space volume is conserved along geodesics \cite{Bertschinger:1993xt}, and number of particles is also conserved, so is the phase space distribution function, namely
\begin{align}
 \frac{{\cal D}}{{\cal D}\lambda}f=\left(\frac{dx^\mu}{d\lambda}\frac{\partial}{\partial x^\mu}+\frac{dp_\mu}{d\lambda}\frac{\partial}{\partial p_\mu}\right)f=0\,,
 \end{align} 
 where $\lambda$ is the affine parameter of the geodesic and ${\cal D}/{\cal D}\lambda$ refers to the Liouville operator. Note that we used $p_\mu\equiv g_{\mu\nu}{dx^\mu}/{d\lambda}$ because the phase space is strictly speaking defined by the covariant conjugate momentum in the tangent space \cite{Ma:1995ey} (see also App.~\ref{app:appendix}). 

 Expanding the collisionless Boltzmann equation in terms of the local inertial frame variables one obtains
\begin{align}\label{eq:Boltzmannlocal}
\left[P^A \tensor{e}{_A^\mu}\frac{\partial }{\partial x^\mu}+\frac{dP_C}{d\lambda}\frac{\partial }{\partial P_C}\right]f=\left[P^A \tensor{e}{_A^\mu}\frac{\partial }{\partial x^\mu}+\tensor{\omega}{_C^B_A}P^AP_B\frac{\partial }{\partial P_C}\right]f=0\,,
\end{align}
where in the last step we used that in the local inertial frame the geodesic equation reads \cite{Saito:2014bxa}
\begin{align}\label{eq:geodesicslocal}
\frac{dP_C}{d\lambda}+\tensor{\omega}{_C^B_A}P^AP_B=0\quad{\rm with}\quad \tensor{\omega}{_C^B_A}P^AP_B=2\tensor{e}{_C^\mu}\tensor{e}{_A^\nu}\partial_{[\nu}\tensor{e}{^B_{\mu]}}P^AP_B\,.
\end{align}
For the Boltzmann equation using the coordinate momentum $p_\mu$ see App.~\ref{app:appendix}. Note that in the last term of Eq.~\eqref{eq:geodesicslocal}, we used that the spin connection in metric spacetimes is given by $\tensor{\omega}{_C^B_A}=\tensor{e}{_C^\nu}\tensor{e}{_A^\mu}\nabla_\mu \tensor{e}{^B_\nu}$, which can also be expressed as $\tensor{\omega}{_C^B_A}=\tensor{e}{^B^\nu}\tensor{e}{_C^\mu}\partial_{[\nu} \tensor{e}{_A_{\mu]}}+\tensor{e}{^B^\nu}\tensor{e}{_A^\mu}\partial_{[\nu} \tensor{e}{_C_{\mu]}}+\tensor{e}{_C^\mu}\tensor{e}{_A^\nu}\partial_{[\nu} \tensor{e}{^B_{\mu]}}$. We used the latter expression  to simplify the contraction $\tensor{\omega}{_C^B_A}P^AP_B$. Note that we consistently used normalized symmetrization, that is, the brackets in the indices carry an additional factor $1/2$ when expanded. We are ready to study linear cosmological perturbations.

\subsection{Linear perturbations\label{sec:linear_perturbations}}

Consider a general perturbed flat FLRW metric in the 3+1 decomposition, given by
\begin{align}\label{eq:metric}
d s^2= g_{\mu\nu}dx^\mu dx^\nu=a^2(\eta)\left(-N^2d\eta^2+H_{ij}(dx^i+N^id\eta)(dx^j+N^jd\eta)\right)\,,
\end{align}
where $a$ is the scale factor, $\eta$ the conformal time, $N$ the lapse, $N^i$ the shift vector and $H_{ij}$ the induced spatial metric. We now fully fix the explicit form of the vierbeins in terms of the metric variables following Refs.~\cite{Naruko:2013aaa,Saito:2014bxa}. To do that, we revisit first the explicit form of the vector $n_\mu$ specifying the time-slice and the induced spatial metric in terms of ADM variables.

First, we note that in the ADM decomposition $n_\mu=a(-N,\vec{0})$, or equivalently, $n_\mu dx^\mu=-aNd\eta$. It then follows that $n^\mu=N^{-1}(1,-N^i)$. In terms of $n_\mu$, the spacetime metric is decomposed as $g_{\mu\nu}=n_\mu n_\mu+a^2H_{\mu\nu}$ where $H_{00}=H_{ij}N^iN^j$ and $H_{0i}=H_{ij}N^i$. However, in its inverse form, that is $g^{\mu\nu}=n^\mu n^\mu+a^{-2}H^{\mu\nu}$, we have that $H^{00}=H^{0i}=0$. Thus, the inverse metric is more convenient to solve for the vierbeins, since we have that
\begin{align}
g^{\mu\nu}=\tensor{e}{_{A}^\mu}\tensor{e}{_{B}^\mu}\eta_{AB}=-\tensor{e}{_{A_0}^\mu}\tensor{e}{_{A_0}^\nu}+\delta^{{A_i}{A_j}}\tensor{e}{_{A_i}^\mu}\tensor{e}{_{A_j}^\nu}\,.
\end{align}
Thus, if we choose $\tensor{e}{_{A_0}^\mu}=n^\mu$ and, by orthogonality, we have $\tensor{e}{_{A_i}^0}=0$, the remaining compatible vierbeins must satisfy
\begin{align}
 \delta^{{A_i}{A_j}}\tensor{e}{_{A_i}^i}\tensor{e}{_{A_j}^j}=a^{-2}H^{ij}\,.
\end{align}

To solve for the vierbeins, we further split the spatial metric as
\begin{align}\label{eq:Hij}
H_{ij}=e^{2\psi}(e^{Y})_{ij}\,,
\end{align}
where $\det H=e^{6\psi}$ and, in Cartesian coordinates, $\det e^Y=1$. It also follows that $(e^{-Y})^{ij}\partial_\eta (e^{Y})_{ij}=\partial_\eta(\det e^Y)=0$, so that $\delta^{ij}Y_{ij}=0$ (see, e.g., Ref.~\cite{Domenech:2017ems}). Negleting vector modes, we further decompose $Y_{ij}$ into the traceless (scalar) and transverse-tranceless (tensor) part, namely $Y_{ij}=2D_{ij}E+h_{ij}$, where $D_{ij}=\partial_i\partial_j-\delta_{ij}\Delta/3$ is the traceless second derivative. Using the properties of exponential matrices, one can show that the vierbeins are given by \cite{Naruko:2013aaa,Saito:2014bxa}
\begin{align}\label{eq:vierbeinup}
\tensor{e}{_{A_0}^\mu}=n^\mu\quad{\rm and}\quad \tensor{e}{_{A_i}^\mu}=\tensor{\delta}{_{A_i}^k}\frac{e^{-\psi}}{a}\delta^{\mu i}(e^{-\frac{1}{2}Y})_{ki}\,.
\end{align}
The vierbein $\tensor{e}{^{A}_\mu}$ follows from the contraction with the metric, that is $\tensor{e}{^{A}_\mu}=g_{\mu\nu}\eta^{{A}{B}}\tensor{e}{_{B}^\mu}$. For completeness, we write them explicitly below,
\begin{align}\label{eq:vierbeindown}
\tensor{e}{^{A_0}_\mu}=-n_\mu\quad{\rm and}\quad\tensor{e}{^{A_i}_\mu}=\delta^{A_ik}\,ae^\psi(e^{\frac{1}{2}Y})_{ki}\left(N^i\delta_\mu^0+\delta_\mu^i\right)\,.
\end{align}
We proceed to derive the perturbed Boltzmann equation and energy-momentum tensor, separately.

\subsubsection{Perturbed Boltzmann equation}

Since we know the vierbeins up to linear order, we can expand the Boltzmann equation in terms of the local momentum. Before Taylor expanding, it is convenient to work in a conformally related frame \cite{Ma:1995ey,Bruni:1996im}. Namely, we define a conformally related metric and vierbeins via
\begin{align}
g_{\mu\nu}=a^2\breve g_{\mu\nu}=a^2(\tensor{E}{_{A}^\mu}\tensor{E}{_{B}^\mu}\eta_{AB})\,,
\end{align}
where
\begin{align}\label{eq:newtetrad}
\tensor{ e}{_{A}^\mu}=a\tensor{E}{_{A}^\mu}\quad{\rm and}\quad \tensor{ e}{^{A}_\mu}=\frac{1}{a}\tensor{ E}{^{A}_\mu}\,.
\end{align}
We define the non-coordinate components of the particle momentum in the conformally related tetrad as ${\mathbf P}=P^A\mathbf{e}_{(A)}=Q^A\mathbf{E}_{(A)}$. From this and Eq.~\eqref{eq:newtetrad}, we infer that
\begin{align}\label{eq:conformallocalP}
Q^A=a P^A\quad{\rm and}\quad Q_A=\frac{1}{a}P_A\,.
\end{align}
As we shall see, $Q^{A_0}=aP^{A_0}=a^2Np^0$ corresponds, at the background level, to the comoving momentum of the particle. Namely, since $P^{A_0}\propto 1/a$ (or, in conformal coordinates, $p^0\propto 1/a^2$) we have at the background that $Q^{A_0}={\rm constant}$.

In the new local conformal frame, the geodesic equation, after some calculations, reads
\begin{align}
\frac{dQ_C}{d\Lambda}+\tensor{\Omega}{_C^B_A}Q^AQ_B=0\quad{\rm with}\quad \tensor{\Omega}{_C^B_A}Q^AQ_B=2\tensor{E}{_C^\mu}\tensor{E}{_A^\nu}\partial_{[\nu}\tensor{E}{^B_{\mu]}}Q^A Q_B\,,
\end{align}
where we redefined the affine parameter via $d \lambda=a^2 d\Lambda$. Note that $\tensor{\Omega}{_C^B_A}$ starts at linear order in perturbation theory, which further confirms that $Q^{A_0}={\rm constant}$ at the background. As shown in App.~\ref{app:appendix}, a similar calculations follows in terms of spacetime components. Lastly, the Boltzmann equation \eqref{eq:Boltzmannlocal} in terms of the conformal local momentum \eqref{eq:conformallocalP} reduces to
\begin{align}\label{eq:Boltzmannlocal}
&\left[Q^A \tensor{E}{_A^\mu}\frac{\partial }{\partial x^\mu}-\tensor{\Omega}{_C^B_A}
Q^AQ_B\frac{\partial }{\partial Q_C}\right]f=\nonumber\\&
\quad \quad Q^{A_0}\left[E_{A_0}^\mu\frac{\partial }{\partial x^\mu}+\hat P^{A_i}\tensor{E}{_{A_i}^\mu}\frac{\partial }{\partial x^\mu}-\frac{
Q^AQ_B}{(Q^{A_0})^2}\left(\tensor{\Omega}{_{A_0}^B_A}\,Q_{A_0}\frac{\partial }{\partial Q_{A_0}}+\tensor{\Omega}{_{A_i}^B_A}\frac{\partial }{\partial \hat P_{A_i}}\right)\right]f=0\,,
\end{align}
where in the last step we explicitly introduced $Q_{A_0}$ and $\hat P_{A_i}$ defined in Eq.~\eqref{eq:Phatai}. 

We now expand up to linear level in perturbation theory. First, we find at the background that
\begin{align}\label{eq:backgroundf}
\left[\frac{\partial }{\partial \eta}+\hat P^{i}\frac{\partial }{\partial x^i}\right]f=0\,,
\end{align}
where we defined $\hat P^i\equiv \hat P^{A_i}\tensor{\delta}{_{A_i}^i}$. From Eq.~\eqref{eq:backgroundf}, we conclude that $f$ can only depend on $Q_{A_0}=-Q^{A_0}$ since it is time independent at the background level. Namely, we have that $f=f(Q^{A_0})$ at the background. At linear level, we perturb the distribution function as
\begin{align}\label{eq:fsplitboltzmann}
f(x^\alpha,Q^{A_0},\hat P^{A_i})=f(Q^{A_0})+\delta f(x^\alpha,Q^{A_0},\hat P^{A_i})\,.
\end{align}
Then, expanding the Boltzmann equation up to linear order we obtain
\begin{align}\label{eq:localboltzmann}
\left[\frac{\partial}{\partial \eta}+\hat P^{i}\frac{\partial }{\partial x^i}\right]\delta f(x^\alpha,Q^{A_0},\hat P^{A_i})+\tensor{\Omega}{_{A_0}^B_A}\frac{
Q^AQ_B}{(Q^{A_0})^2}\,Q^{A_0}\frac{\partial f(Q^{A_0}) }{\partial Q^{A_0}}=0\,,
\end{align}
where note that we used $Q^{A_0}$ for convenience. 
After some calculations, one can check that \cite{Saito:2014bxa}
\begin{align}
\tensor{\Omega}{_{A_0}^B_A}\frac{
Q^AQ_B}{(Q^{A_0})^2}=-\hat P^i\partial_i\phi+\hat P^i\hat P^j\left(\partial_i\partial_j\beta - \psi'\delta_{ij}-\frac{1}{2}Y_{ij}'\right)\,,
\end{align}
where we used that $N=e^\phi$ and $N_i=\partial_i\beta$.

To simplify the form of the Boltzmann equation, it is standard practice to introduce a normalized distribution function fluctuation given by
\begin{align}\label{eq:gammadef}
\Gamma(x^\alpha,Q^{A_0},\hat P^{A_i})\equiv \frac{\delta f(x^\alpha,Q^{A_0},\hat P^{A_i})}{-Q^{A_0}\frac{\partial f(Q^{A_0}) }{\partial Q^{A_0}}}\,,
\end{align}
not to be confused with the Christoffel symbols. We used $\Gamma$ following the notation of Ref.~\cite{Contaldi:2016koz} for the graviton gas. Note that, in CMB calculations where the photons follow a thermal spectrum, the standard convention is to use $\Gamma\to\Theta=\delta T/T$ instead \cite{Dodelson:2003ft}. In terms of $\Gamma$, the Boltzmann equation \eqref{eq:localboltzmann} becomes
\begin{align}\label{eq:gammaequation}
\Gamma'+\hat P^{i}\partial_i\Gamma+\hat P^i\partial_i\phi+{\cal R}'-\hat P^i\hat P^j\partial_i\partial_j\sigma+\hat P^i\hat P^j h'_{ij}=0\,,
\end{align}
where we introduced the curvature and shear perturbations ${\cal R}=\psi-\frac{1}{3}\Delta E$ and $\sigma=\beta - E'$ respectively. Notably, Eq.~\eqref{eq:localboltzmann} is independent of $Q^{A_0}$, or in other words, it is manifestly frequency independent. Thus, any frequency dependence in $\Gamma$ must only come from initial conditions. As we later prove, this expression is also gauge invariant. We checked that our results coincides with those in Refs.~\cite{SasakiCPTlectures,Naruko:2013aaa}.

\subsubsection{Perturbed energy-momentum tensor}

We have derived above the perturbed Boltzmann equation. However, Einstein equations depend on the energy-momentum tensor of the particle gas and so do the adiabatic initial conditions. Thus, we must explicitly compute the components of the energy-momentum tensor in terms of spacetime components. Note that we first keep the local momentum inside the momentum integrals for convenience (and because it is more appropriate from the point of view of the definition of momentum in the tangent space). At the end of this section, we also show the result for energy density in terms of spacetime momentum components.

If we keep the local momenta inside the momentum integral (as in Eq.~\eqref{eq:TAB}), we find that Eq.~\eqref{eq:Tmunu}, with one index up, reads
\begin{align}
\tensor{T}{^{\mu}_{\nu}}=\tensor{e}{_A^\mu}\tensor{e}{^B_\nu}\tensor{T}{^{A}_{B}}=\tensor{e}{_A^\mu}\tensor{e}{^B_\nu}\int dP^{A_0}d\Omega \,(P^{A_0})^2\,\frac{P^AP_B}{|P_{A_0}|}f(x^\mu,P_I)\,,
\end{align}
where we used that $dP_{A_1}dP_{A_2}dP_{A_3}=(P^{A_0})^2dP^{A_0}d\Omega$ (see Eq.~\eqref{eq:Phatai}), with $d\Omega$ being the solid angle in the local momentum space.
A straightforward expansion of the spacetime components yields
\begin{align}
\tensor{T}{^{0}_{0}}&= \tensor{e}{_{A_0}^0}\tensor{e}{^{A_0}_0}\,\tensor{T}{^{A_0}_{A_0}}+\tensor{e}{_{A_0}^0}\tensor{e}{^{A_i}_0}\,\tensor{T}{^{A_0}_{A_i}}\,,\\
\tensor{T}{^{0}_{i}}&= \tensor{e}{_{A_0}^0}\tensor{e}{^{A_i}_i}\,\tensor{T}{^{A_0}_{A_i}}\,,\\
\tensor{T}{^{i}_{j}}&= \tensor{e}{_{A_0}^i}\tensor{e}{^{A_j}_j}\,\tensor{T}{^{A_0}_{A_j}}+\tensor{e}{_{A_i}^i}\tensor{e}{^{A_j}_j}\,\tensor{T}{^{A_j}_{A_i}}\,.
\end{align}

We now use the form of the perturbed vierbeins, Eqs.~\eqref{eq:vierbeinup} and \eqref{eq:vierbeindown}, as well as the fact that the background Boltzmann equation yields $f(x^\alpha,Q^{A_0},\hat P^{A_i})=f(Q^{A_0})+\delta f(x^\alpha,Q^{A_0},\hat P^{A_i})$ (see discussion around Eq.~\eqref{eq:fsplitboltzmann}). Using also that $\int d\Omega \,\hat P^{A_i}=0$ \cite{Ma:1995ey}, we obtain at linear order that
\begin{align}\label{eq:T00}
\tensor{T}{^{0}_{0}}&=- a^{-4}\int d\Omega \,dQ^{A_0} \,(Q^{A_0})^3\,(f(Q^{A_0})+\delta f(x^\alpha,Q^{A_0},\hat P^{A_i}))\,,\\
\tensor{T}{^{0}_{i}}&=a^{-4}\int d\Omega \,dQ^{A_0}\,(Q^{A_0})^3\,\hat P_{i}\,\delta f(x^\alpha,Q^{A_0},\hat P^{A_i})\,,\\
\tensor{T}{^{i}_{j}}&= a^{-4}\int d\Omega \,dQ^{A_0} \,(Q^{A_0})^3\,\hat P^{i}\hat P_{j}\,(f(Q^{A_0})+\delta f(x^\alpha,Q^{A_0},\hat P^{A_i}))\,.
\end{align}
 We identify the background and perturbed energy density of the particle gas comparing Eq.~\eqref{eq:T00} with the result of a perfect fluid, that is $\tensor{T}{^0_0}=-(\rho+\delta\rho)$. This yields
\begin{align}\label{eq:rhoanddeltarho}
\rho=a^{-4}\int d\Omega \,dQ^{A_0} \,(Q^{A_0})^3\,f(Q^{A_0})\quad{\rm and}\quad \delta\rho=a^{-4}\int d\Omega \,dQ^{A_0} \,(Q^{A_0})^3\,\delta f(x^\alpha,Q^{A_0},\hat P^{A_i})\,.
\end{align}
Let us emphasize that inside the momentum integral we have the local conformal momentum $Q^{A_0}$. 

Before dealing with the gauge invariance, it is interesting to note that, if we use the spacetime components of the 4-momentum, $p_\mu$, when computing $\tensor{T}{^{0}_{0}}$, we get from Eq.~\eqref{eq:Tmunu} that
\begin{align}\label{eq:T00spacetime}
\tensor{T}{^{0}_{0}}=-\frac{1}{a^4N}e^{-3\psi}\int {d^3p}\,p^0 \,f(x^\sigma,p_\beta)\,.
\end{align} 
Note that, by definition of phase space, the integral depends on the covariant spatial momentum, namely $d^3p=dp_x dp_y dp_z$ in Cartesian coordinates. This is why one obtains a factor $1/a^4$ in Eq.~\eqref{eq:T00spacetime} when using conformal time. Thus, for consistency, we must express $p^0$ in terms of the covariant spatial momentum $p_i$ too.

For simplicity and comparison with the literature, let us fix the Newton gauge and neglect tensor perturbations, that is we set $\beta=E=h_{ij}=0$, $\psi=\Psi$ and $\phi=\Phi$ in Eq.~\eqref{eq:metric}. In that case, the null condition, that is $g_{\mu\nu}p^\mu p^\nu=-N^2(p^0)^2+H_{ij}p^ip^j=0$, yields
\begin{align}\label{eq:p0solved}
p^0=e^{-(\Phi+\Psi)}|\vec{p}|\quad {\rm where}\quad |\vec{p}|^2\equiv \delta^{ij}p_ip_j\,,
\end{align}
and we further used that $N=e^\Phi$  and $p_i=H_{ij}p^j$, with $H_{ij}$ given by Eq.~\eqref{eq:Hij}.
Inserting Eq.~\eqref{eq:p0solved} into Eq.~\eqref{eq:T00spacetime}, we arrive at
\begin{align}\label{eq:T00spacetimemomentum}
\tensor{T}{^{0}_{0}}=-\frac{1}{a^4}e^{-2\Phi-4\Psi}\int {d^3p}|\vec{p}| f(x^\sigma,p_\beta)\,.
\end{align}
If we now expand Eq.~\eqref{eq:T00spacetimemomentum} at linear order, and use that $\tensor{T}{^{0}_{0}}=-(\rho+\delta\rho)$, we identify instead
\begin{align}\label{eq:rhononlocal}
\rho=\frac{1}{a^4}\int {d^3p}|\vec{p}| f(|\vec{p}|)\quad {\rm and}\quad \delta\rho=-(2\Phi+4\Psi)\rho+\frac{1}{a^4}\int {d^3p}|\vec{p}| \delta f(x^\sigma,p_\beta)\,.
\end{align}
We note that the factor  $-(2\Phi+4\Psi)\rho$ is the same as the one found in Refs.~\cite{ValbusaDallArmi:2023nqn,ValbusaDallArmi:2024hwm,Mierna:2024pkh} (after properly mapping different notations, namely $\Phi_{\rm here}\to \Psi_{\cite{ValbusaDallArmi:2024hwm}}$ and $\Psi_{\rm here}\to- \Phi_{\cite{ValbusaDallArmi:2024hwm}}$) when studying fluctuations of the GW energy density in cosmology from the Isaacson prescription. It is precisely this term which appears to yield ``non-adiabatic'' initial conditions for GW fluctuations \cite{ValbusaDallArmi:2023nqn,ValbusaDallArmi:2024hwm}.

But, as we have shown in Eq.~\eqref{eq:rhoanddeltarho}, the term $-(2\Phi+4\Psi)\rho$ is absent when one uses the local momentum in the integral, leading to a clearer identification between $\delta\rho$ and $\delta f$ \eqref{eq:rhoanddeltarho}. Note that the local momentum is also the one more appropriately defined in the tangent space and, furthermore, it significantly simplifies the Boltzmann equation (see Eq.~\eqref{eq:gammaequation}). Thus, it is more convenient to define adiabatic initial conditions using the local momentum and so Eq.~\eqref{eq:rhononlocal}. We note that our results are valid as long as we can describe the GW background in terms of a gas of gravitons. However, this approximation breaks down for sufficiently low, super-Hubble momenta. We leave a detailed study beyond the graviton gas approximation for future work.

\subsection{Gauge invariance\label{sec:gauge_invariance}}

We now show that the linear Boltzmann equation \eqref{eq:gammaequation} is gauge invariant. This serves as a consistency check of our previous results and will be helpful for the later discussion on the adiabatic initial conditions. We first study the gauge transformation of all relevant quantities and later turn show the gauge invariance of the Boltzmann equation.

Consider the coordinate transformation given by
\begin{align}
\tilde x^\mu=x^\mu+\xi^\mu\quad{\rm with}\quad \xi^\mu=(T,\partial^i L)\,.
\end{align}
Under such change of coordinates, we find that the metric transforms as
\begin{align}
\tilde g_{\mu\nu}= g_{\mu\nu}-\xi^\alpha\partial_\alpha g_{\mu\nu}-2g_{\alpha(\mu}\partial_{\nu)}\xi^{\alpha}\,,
\end{align}
and so it follows that
\begin{align}\label{eq:gaugephi}
\tilde\phi= \phi-{\cal H}T-T'\quad,\quad
\tilde\beta= \beta+T-L'\quad,\quad
\tilde\psi= \psi-{\cal H}T-\frac{1}{3}\Delta L\quad,\quad
\tilde E= E-L\,.
\end{align}
Tensor modes are trivially gauge invariant, that is $\tilde h_{ij}= h_{ij}$. For the canonical momentum, we have by definition that
\begin{align}\label{eq:pmu}
\tilde p^\mu=\frac{d\tilde x^\mu}{d\lambda}= p^\mu+p^\alpha\partial_\alpha\xi^\mu\,.
\end{align}

With the above transformation rules we compute the change in the local momenta $P^A$ from their expression in terms of metric variables and the 4-momentum, namely we use that
\begin{align}
P^{A_0}&=e^{A_0}_\mu p^\mu=a(1+\phi)p^0\,,\\
P^{A_i}&=e^{A_i}_\mu p^\mu=ap^0\partial_i\beta \tensor{\delta}{^{A_i}^i}+ap^i\left((1+\psi)\delta_{ij}+\frac{1}{2}Y_{ij}\right)\tensor{\delta}{^{A_i}^j}\,,
\end{align}
where we used Eq.~\eqref{eq:vierbeindown} for the vierbeins.
Using Eqs.~\eqref{eq:gaugephi} and \eqref{eq:pmu}, we conclude that
\begin{align}
\tilde P^{A_0}=  P^{A_0}+P^{A_i}\tensor{\delta}{_{A_i}^i}\partial_iT\quad{\rm and}\quad
\tilde P^{A_i}=  P^{A_i}+P^{A_0} \tensor{\delta}{^{A_i}^i}\partial_iT\,,
\end{align}
At this point, it is interesting to note that the transformation rules for the local momentum follow that of a local Lorentz transformation \cite{Naruko:2013aaa,Saito:2014bxa}. Indeed, it is easy to see that
\begin{align}
\tilde P^A= \tensor{\Lambda}{^A_B}P^B\,, 
\end{align}
with
\begin{align}
\tensor{\Lambda}{^{A_0}_{A_0}}=1\quad,\quad\tensor{\Lambda}{^{A_0}_{A_i}}=\tensor{\Lambda}{^{A_i}_{A_0}}=\tensor{\delta}{_{A_i}^i}\partial_iT\quad,\quad \tensor{\Lambda}{^{A_i}_{A_j}}=\tensor{\delta}{^{A_i}_{A_j}}\,.
\end{align}
In other words, if we impose the same requirements to the tetrads before and after the gauge transformation, namely we require that the time-like tetrad points along the proper time direction, we find that
\begin{align}
\mathbf{\tilde e}_{(A)}= \tensor{\Lambda}{_{A}^{B}}\mathbf{e}_{(B)}\,,
\end{align}
so that the change of coordinates leads to an additional a Lorentz transformation of the tetrads.

We proceed to show the gauge invariance of the Boltzmann equation. Since the distribution function at the background depends on $Q^{A_0}$ \eqref{eq:fsplitboltzmann}, we need the gauge transformation of $Q^{A_0}$. From $Q^{A_0}=aP^{A_0}$, it follows that
\begin{align}\label{eq:tildeQA0}
\tilde Q^{A_0}&=  Q^{A_0}\left(1+{\cal H}T+ \hat P^{i}\partial_iT\right)\,.
\end{align}
Furthermore, using that the distribution function is a scalar, we have that
\begin{align}
\tilde f(\tilde Q^{A_0})+\delta\tilde f= f( Q^{A_0})+\delta f\,.
\end{align}
In this way, we obtain that
\begin{align}\label{eq:deltaftransform}
\delta \tilde f= \delta f - ({\cal H}T+\hat P^i\partial_i T)\,Q^{A_0}\frac{\partial f}{\partial Q^{A_0}}\,,
\end{align}
which from Eq.~\eqref{eq:gammadef} yields
\begin{align}\label{eq:Gamma_transform}
\tilde\Gamma=  \Gamma +{\cal H}T+\hat P^i\partial_i T\,.
\end{align}

Using Eqs.~\eqref{eq:tildeQA0} and \eqref{eq:gaugephi}, it is not difficult to convince oneself that
\begin{align}
&\tilde\Gamma'+\hat P^{i}\partial_i\tilde\Gamma+\hat P^i\partial_i\tilde\phi+\tilde{\cal R}'-\hat P^i\hat P^j\partial_i\partial_j\tilde \sigma+\hat P^i\hat P^j \tilde h'_{ij}=\nonumber\\&\qquad\qquad
\Gamma'+\hat P^{i}\partial_i\Gamma+\hat P^i\partial_i\phi+{\cal R}'-\hat P^i\hat P^j\partial_i\partial_j\sigma+\hat P^i\hat P^j h'_{ij}=0\,.
\end{align}
With this we conclude that our formulation in terms of the local tetrad is consistent and that we understand the gauge transformation rules of each variable, specially the local momentum and distribution function. For a proof using the spacetime momentum see Ref.~\cite{SasakiCPTlectures} and for a proof of the gauge invariance at second order in perturbation theory see Ref.~\cite{Naruko:2013aaa}.

\section{Classification of Adiabatic initial conditions \label{sec:adiabatic_conditions}}

In the previous section we reviewed the formulation of a gas of particles in a FLRW background at linear order in perturbation theory. We have derived the gauge transformation for each variable and checked that the Boltzmann equation is gauge invariant. We are now ready to study in detail the adiabatic initial conditions for a relativistic gas on super-Hubble scales. We do so first by requiring the absence of isocurvature in Sec.~\ref{sec:common_uniform} and by the separate universe approach in Sec.~\ref{sec:separate_universe}. The aim of this section is to provide a complete classification of possible adiabatic initial conditions for a gas of relativistic particles.

\subsection{Common uniform density slice\label{sec:common_uniform}}

Consider that we have $X$ fluids in the universe, each one with its energy density and pressure $\rho_X$ and $P_X$, and its fluctuations, say $\delta\rho_X$. The total curvature perturbation on uniform density slices in the Universe is defined by \cite{Langlois:2003fq,Malik:2004tf}
\begin{align}\label{eq:zeta}
\zeta={\cal R} -\frac{1}{3}\frac{\delta\rho}{\rho+P}\,,
\end{align}
where $\delta\rho=\sum_X\delta\rho_X$. The same holds for $\rho$ and $P$. Recall that ${\cal R}$ is defined below Eq.~\eqref{eq:gammaequation}. One may also define individual curvature perturbations for each fluid \cite{Malik:2004tf}, namely
\begin{align}\label{eq:zetaX}
\zeta_X={\cal R} -\frac{1}{3}\frac{\delta\rho_X}{\rho_X+P_X}\,.
\end{align}
The individual curvature perturbations $\zeta_X$ is useful to introduce the notion of isocurvature between fluid $X$ and fluid $Y$ \cite{Malik:2004tf}, i.e., $S_{XY}=3(\zeta_X-\zeta_Y)$. Isocurvature also quantifies relative number density fluctuations between fluids \cite{Langlois:2003fq}.

Adiabatic initial conditions are defined as the absence of isocurvature, namely $S_{XY}=0$, $\forall X,Y$, from which it follows that
\begin{align}
\zeta=\zeta_X \,;\, \forall X\,.
\end{align}
At the same time, from Eqs.~\eqref{eq:zeta} and \eqref{eq:zetaX}, adiabatic initial conditions require that there is common time-slice where the energy density of each and every fluid is spatially homogeneous, namely that
\begin{align}\label{eq:densityzero}
\delta\rho=0\quad{\rm and}\quad\delta\rho_X=0\,;\, \forall X\,.
\end{align}
This is our definition of adiabatic initial conditions on super-Hubble scales. 

Most commonly, however, one works in the Newton slice for convenience, which corresponds to a time-slice with no shear. We can go from the uniform  total density time-slice to the Newton slice with a time reparametrization given by $\tilde \eta=\eta+T(t,x)$. The energy density fluctuations for each fluid after such gauge transformation read \cite{Malik:2008im}
\begin{align}\label{eq:deltarhoX}
\delta\tilde \rho_X = \delta\rho_X - \rho_X' T\,.
\end{align}
The same applies to the total energy density fluctuation, that is $\delta\tilde \rho= \delta\rho - \rho' T$. Now, if we say that tilde quantities are evaluated in the  uniform total density slice, denoted with a superscript ``$\rm ud$'', we have that
\begin{align}\label{eq:deltarhozero}
\delta \rho^{\rm ud}= \delta\rho^{\rm N} - \rho' T^{\rm N}=0\,,
\end{align}
where ``$\rm N$'' refers to Newton slice. Solving Eq.~\eqref{eq:deltarhozero} for  $T^{\rm N}$, we conclude that the time gauge parameter moving from the uniform total density to the Newton slice is given by
\begin{align}\label{eq:tNsolution}
T^{\rm N}=-\frac{1}{3{\cal H}}\frac{\delta\rho^{\rm N}}{\rho+P}\,,
\end{align}
and is the same for all fluids.

We can write the adiabatic initial conditions in a more standard form, by expressing $T^{\rm N}$ in terms of the Newtonian potential $\Phi$ after using Einstein equations. In particular, the $00$ component of Einstein equations on super-Hubble scales (that is we neglect derivative terms) reads \cite{Malik:2008im}
\begin{align}\label{eq:phisuperhubble}
\Phi({\rm super-Hubble})\approx -\frac{1}{2}\frac{\delta\rho^{\rm N}}{\rho}\,.
\end{align}
Using Eqs.~\eqref{eq:deltarhoX} and \eqref{eq:phisuperhubble}, we find that adiabatic initial conditions imply
\begin{align}
\frac{\delta\rho_X^{\rm N}}{\rho_X}=-3{\cal H}(1+w_X)T^{\rm N}=\frac{1+w_X}{1+w}\frac{\delta\rho^{\rm N}}{\rho+P}=-2\Phi\frac{1+w_X}{1+w} \,,
\end{align}
where we introduced the equation of state $w_X=P_X/\rho_X$. This means that, in the very early, radiation dominated universe, where $w\approx 1/3$, a radiation-like fluid, such as a gas of relativistic particles, with adiabatic initial conditions on super-Hubble scales and in the Newton slice, initially has
\begin{align}\label{eq:standardadiabatic}
\frac{\delta\rho_{\rm rad}^{\rm N}}{\rho_{\rm rad}}=-2\Phi\,.
\end{align}

We can proceed similarly for a gas of massless particles. This time, however, the uniform density slice results in an integral constraint on the distribution function's fluctuations, as it requires $\delta\rho$ in Eq.~\eqref{eq:rhoanddeltarho} to vanish, namely
\begin{align}\label{eq:deltarhouniform}
 \delta\rho^{\rm ud}=a^{-4}\int d\Omega \,dQ^{A_0} \,(Q^{A_0})^3\,\delta f^{\rm ud}(x^\alpha,Q^{A_0},\hat P^{A_i})=0\,.
\end{align}
From Eq.~\eqref{eq:deltarhouniform}, we identify two distinct solutions which we list below.
\begin{enumerate}[label = (\itshape\roman *)]
\item \textit{Strong} adiabatic initial conditions (s.a.), where
\begin{align}\label{eq:strongud}
\delta f^{\rm ud}_{\rm (s.a.)}=0\Rightarrow \Gamma^{\rm ud}_{\rm (s.a.)}=0\,,
\end{align}
and the distribution function's fluctuations vanish independent of momentum.
\item \textit{Weak} adiabatic initial conditions (w.a.), where
\begin{align}\label{eq:weakud}
\int d\Omega \,dQ^{A_0} \,(Q^{A_0})^3\,\delta f^{\rm ud}_{\rm (w.a.)}=0\Rightarrow \int d\Omega \,dQ^{A_0} \,(Q^{A_0})^3\,\left(-Q^{A_0}\frac{\partial f}{\partial Q^{A_0}}\right)\,\Gamma^{\rm ud}_{\rm (w.a.)}=0\,.
\end{align}
In this case, we have that $\delta f^{\rm ud}_{\rm (w.a.)},\Gamma^{\rm ud}_{\rm (w.a.)}\neq0$ and $\delta\rho^{\rm ud}_{\rm (w.a.)}$ only vanishes after integration. Note that there may be many choices of $\delta f^{\rm ud}_{\rm (w.a.)}$ for which the integral vanishes. 
\end{enumerate}
We may also refer to the weak adiabatic initial conditions as \textit{internal isocurvature} because, on super-Hubble scales, one could think of each momentum slice of the distribution function as separate fluids. Thus, condition $(ii)$ can be viewed as compensated energy density fluctuations among the different momentum contributions. For example, if we consider that $\delta f^{\rm ud}(Q^{A_0})$ has only two contributions, say $\delta f^{\rm ud}(Q^{A_0})=\delta f^{\rm ud}_1(Q^{A_0}_1)\delta(\ln(Q^{A_0}/Q^{A_0}_1))+\delta f^{\rm ud}_2(Q^{A_0}_2)\delta(\ln(Q^{A_0}/Q^{A_0}_2))$. Then, the weak adiabatic initial condition imposes $(Q^{A_0}_1)^4\delta f^{\rm ud}_1(Q^{A_0}_1)=-(Q^{A_0}_2)^4\delta f^{\rm ud}_2(Q^{A_0}_2)$. Note that both initial conditions have been considered in the literature. For instance, condition $(i)$ is used in Refs.~\cite{Dimastrogiovanni:2022eir,Malhotra:2022ply} and condition $(ii)$ in Refs.~\cite{Ricciardone:2021kel,Schulze:2023ich}.

Let us see what the different adiabatic initial conditions imply in the Newton gauge. Since we know the gauge transformation of $\delta f$ and $\Gamma$, see Eqs.~\eqref{eq:deltaftransform} and \eqref{eq:Gamma_transform}, we can directly compute the distribution function's fluctuations in the Newton gauge using $T^N$ given by Eqs.~\eqref{eq:tNsolution} and \eqref{eq:phisuperhubble}. Doing so, we find that the possible adiabatic initial conditions in the Newton gauge are:
\begin{enumerate}[label = (\itshape\roman *)]
\item \textit{Strong} adiabatic initial conditions, where
\begin{align}\label{eq:stronggammaN}
\Gamma^{\rm N}_{\rm(s.a.)}=-\frac{1}{2}\Phi\,.
\end{align}
\item \textit{Weak} adiabatic initial conditions (or {internal isocurvature}), where
\begin{align}\label{eq:weakgammaN}
\int d\Omega \,dQ^{A_0} \,(Q^{A_0})^3\,\left(-Q^{A_0}\frac{\partial f}{\partial Q^{A_0}}\right)\,\Gamma^{\rm N}_{\rm(w.a.)}=\frac{\Phi}{2}\int d\Omega \,dQ^{A_0} \,(Q^{A_0})^4\,\frac{\partial f}{\partial Q^{A_0}}\,\,.
\end{align}
\end{enumerate}

For the weak adiabatic initial conditions in the Newton gauge, the notion of internal isocurvature appears as follows. First, we note that
\begin{align}\label{eq:relationtogammaiso}
\Gamma^{\rm N}_{\rm(w.a.)}=\Gamma^{\rm N}_{\rm(s.a.)}+\Gamma_{\rm iso}^{\rm N}\,,
\end{align}
where the subscript ``iso'' refers to internal isocurvature and $\Gamma_{\rm iso}^{\rm N}$ satisfies
\begin{align}\label{eq:weakgammaiso}
\int d\Omega \,dQ^{A_0} \,(Q^{A_0})^3\,\left(-Q^{A_0}\frac{\partial f}{\partial Q^{A_0}}\right)\,\Gamma_{\rm iso}^{\rm N}=\int d\Omega \,dQ^{A_0} \,(Q^{A_0})^3\,\delta f_{\rm iso}^{\rm N}=0\,.
\end{align}
As before, there may be many choices of $\Gamma^{\rm N}_{\rm iso}$ for which the integral vanishes. Note that $\delta f_{\rm iso}^{\rm N}=\delta f_{\rm (w.a.)}^{\rm ud}$ and, therefore, $\delta f_{\rm iso}$ satisfying Eq.~\eqref{eq:weakgammaN} is gauge invariant. This is also clear from its definition, Eq.~\eqref{eq:relationtogammaiso}, that can be recasted as $\Gamma_{\rm iso}^{\rm N}=\Gamma^{\rm N}_{\rm(w.a.)}-\Gamma^{\rm N}_{\rm(s.a.)}$. Since it is a difference, it must be gauge invariant. From now on, we simply use $\delta f_{\rm iso}$ and $\Gamma_{\rm iso}$ for the internal isocurvature. Since the Boltzmann equation for $\Gamma$ \eqref{eq:gammaequation} is independent of $Q^{A_0}$, any internal isocurvature is preserved throughout the evolution. Namely, initial internal isocurvature remains constant.

For clarity, it should be note that the strong adiabatic initial condition \eqref{eq:stronggammaN} coincides with the standard one for CMB temperature fluctuations, where one has $\Gamma^{\rm N}_{(\rm s.a.)}=\Theta=\delta T/T$ \cite{Dodelson:2003ft}. Now, if we compute $\delta\rho^{\rm N}_{(\rm s.a.)}$ from Eq.~\eqref{eq:rhoanddeltarho}, we find that
\begin{align}\label{eq:deltarhonewtonian0}
\delta\rho^{\rm N}_{(\rm s.a.)}&=a^{-4}\int d\Omega \,dQ^{A_0} \,(Q^{A_0})^3\,\left(-Q^{A_0}\frac{\partial f}{\partial Q^{A_0}}\right) \Gamma^{\rm N}_{\rm (s.a.)}\nonumber\\&\qquad\qquad\qquad=-2\Phi\rho+\frac{4\pi}{a^4}\frac{\Phi}{2}\left[(Q^{A_0})^4 f(Q^{A_0})\right]^{\infty}_{0}=-2\Phi\rho\,.
\end{align}
where we used Eq.~\eqref{eq:stronggammaN} for $\Gamma^{\rm N}_{\rm (s.a.)}$ and Eq.~\eqref{eq:rhoanddeltarho} for $\rho$. In the last step, we assumed that the distribution function falls off fast enough so that the boundary terms vanish. We note that one needs such conditions to reliably compute the energy density in both the low and high momentum limits. In general, the same result for $\delta\rho$ is obtained for the weak adiabatic initial condition, as they only differ by internal isocurvature fluctuations, which trivially vanish after integration.

As we argued below Eq.~\eqref{eq:weakgammaiso}, the weak adiabatic condition does not fully specify the functional form of $\Gamma^{\rm N}_{\rm (w.a.)}$, which is needed for the Boltzmann equation, because of the freedom in the functional form of the internal isocurvature freedom. One such possibility is given by Refs.~\cite{Ricciardone:2021kel,Schulze:2023ich}, which in our notation corresponds to
\begin{align}
\Gamma^{\rm N}_{\rm (w.a.)\star}=-{2\Phi}\left(-{\frac{\partial\ln f}{\partial \ln Q^{A_0}}}\right)^{-1}\quad{\rm and}\quad\delta f_{\rm (w.a.)\star}= -2 \Phi f\,,
\end{align}
where the star in the subscript denotes that this is a special choice of weak adiabatic initial conditions.
Using Eq.~\eqref{eq:relationtogammaiso}, this choice corresponds to
\begin{align}
\Gamma_{\rm iso \star}=-\frac{\Phi}{2}\left(-{\frac{\partial\ln f}{\partial \ln Q^{A_0}}}\right)^{-1}\left(4+{\frac{\partial\ln f}{\partial \ln Q^{A_0}}}\right)\,.
\end{align}
The interesting aspect of $\Gamma^{\rm N}_{\rm (w.a.)\star}$ is that the energy density fluctuation in the Newton gauge is given by
\begin{align}\label{eq:deltarhonewtonian0}
\delta\rho^{\rm N}_{(\rm w.a.)\star}&=a^{-4}\int d\Omega \,dQ^{A_0} \,(Q^{A_0})^3\,\left(-Q^{A_0}\frac{\partial f}{\partial Q^{A_0}}\right)\Gamma^{\rm N}_{\rm (w.a.),\star}=-2\Phi\rho\,,
\end{align}
without the need to impose any condition on boundary terms. However, one must still require the boundary terms to vanish to have a well-defined common uniform density slice and absence of integrated isocurvature (see discussion around Eq.~\eqref{eq:densityzero}). Namely, condition Eq.~\eqref{eq:deltarhouniform} requires
\begin{align}\label{eq:isoboundary}
\delta\rho^{\rm ud}_{\rm (w.a.)\star}&=a^{-4}\int d\Omega \,dQ^{A_0} \,(Q^{A_0})^3\,\left(-Q^{A_0}\frac{\partial f}{\partial Q^{A_0}}\right)\,\Gamma_{\rm iso \star}\nonumber\\&=-\frac{4\pi}{a^4}\frac{\Phi}{2}\left[(Q^{A_0})^4 f(Q^{A_0})\right]^{\infty}_{0}=0\,.
\end{align}

One may wonder whether the vanishing of the boundary terms, as in Eqs.~\eqref{eq:deltarhouniform} and \eqref{eq:isoboundary}, is too strong a condition. Indeed, we encounter conceptual problems if we consider an almost scale invariant GW spectrum with low and high momentum cut-offs, as naively expected from inflation \cite{Guzzetti:2016mkm}. However, it should be noted that detailed studies of the high momentum tail of the GW spectrum generated during inflation find an exponential decay for large momentum \cite{Negro:2024iwy,Pi:2024kpw}. In the opposite limit,  that is for low momentum, one expects that the Isaacson prescription \cite{Isaacson:1968hbi} stops being valid, as GWs become frozen anisotropies (constant tensor modes) on scales larger than the Hubble radius. And although it is not clear what is the contribution from these modes to the energy density, the general expectation is that it quickly vanishes as the momentum of the graviton vanishes. The main intuition behind is that a constant anisotropy can be removed by a rotation of the spatial coordinates \cite{Bordin:2016ruc,Bartolo:2022wqq}. Although a general investigation is required to make a definitive claim, it is plausible that the boundary terms in \eqref{eq:deltarhouniform} vanish. If so, we have a consistent formulation of adiabatic initial conditions in any gauge. Let us emphasize, though, that both strong and weak adiabatic initial conditions are valid initial conditions and must be determined by studying the concrete  generation mechanism. With this, we conclude the classification of possible adiabatic initial conditions.

\subsection{Separate universe approach \label{sec:separate_universe}}

We end this section with a more heuristic, yet more intuitive, way to derive the adiabatic initial conditions. In the so-called separate universe approach one starts from a homogeneous universe and perform a time translation, say $\eta\to \eta+\delta \eta(\eta,x)$, on that background. Note that this initial assumption should eventually be consistent with the existence of a common uniform density slice, as we did in Sec.~\ref{sec:common_uniform}.

As an example, consider the energy density of a radiation fluid which decays as
\begin{align}
\rho(\eta)=\rho_*(a(\eta)/a_*)^{-4}\,,
\end{align}
where $\rho_*$ and $a_*$ are evaluated at a pivot time $\eta=\eta_*$. If we do the time translation, $\eta\to \eta+\delta \eta(\eta,x)$, we see that
\begin{align}
\rho(\eta)\to \rho(\eta+\delta \eta)=\rho(\eta)+\frac{\partial\rho}{\partial \eta}\delta \eta=\rho(\eta)-4{\cal H}\rho \delta \eta\,.
\end{align}
One then identifies the term proportional to $\delta \eta$ as the adiabatic energy density fluctuation, namely
\begin{align}\label{eq:deltarhooverrho}
\frac{\delta\rho}{\rho}=-4{\cal H}\delta \eta\,.
\end{align}
We can fix the form of $\delta \eta$ in terms of the Newton potential $\Phi$ using again the $00$ component of Einstein equations in the Newton gauge and on super-Hubble scales \cite{Malik:2008im}. This yields
\begin{align}\label{eq:deltaeta}
{\cal H}\delta \eta^{\rm N}=\frac{\Phi}{2}\,.
\end{align}
Inserting Eq.~\eqref{eq:deltaeta} into \eqref{eq:deltarhooverrho}, we recover the adiabatic initial conditions on super-Hubble scales given by Eq.~\eqref{eq:standardadiabatic}. The same procedure is valid for any fluid in the universe and should be valid for a gas of particles.

Let us apply the same logic to the distribution function $f(Q^{A_0})$. Before that, recall that $Q^{A_0}=a(\eta) P^{A_0}$ where $P^{A_0}$ is the local momentum of the particle. To see how the local momentum transforms in the separate universe approach, consider a local Minkowski background metric, that is
\begin{align}
ds^2_{\rm local}=-d\eta^2+\delta_{ij}dx^idx^j=-(\tensor{e}{^{A_0}_0}d\eta)^2+\delta_{{A_i}{A_j}}\tensor{e}{^{A_i}_{i}}\tensor{e}{^{A_j}_{j}}dx^idx^j\,,
\end{align}
where in the last step we introduced the local vierbeins with $\tensor{e}{^{A_0}_0}=1$ and $\tensor{e}{^{A_i}_{i}}=\tensor{\delta}{^{A_i}_i}$ (the other spacetime components equal to zero). Note that the inclusion of the scale factor in the metric would only make calculations more cumbersome but would not change the result. Now, we perform the time translation $\eta\to \eta+\delta \eta(x,\eta)$. This yields at leading order in $\delta\eta$ that
\begin{align}\label{eq:dslocaltransf}
ds^2_{\rm local}\to ds^2_{\rm local}&= -(1+2\delta\eta')d\eta^2+2\partial_i\delta \eta d\eta dx^i+\delta_{ij}dx^idx^j\nonumber\\&=-(\tensor{\tilde e}{^{A_0}_\mu} dx^\mu)^2+\delta_{{A_i}{A_j}}\tensor{\tilde e}{^{A_i}_{i}}\tensor{\tilde e}{^{A_j}_{j}}dx^i dx^j
\nonumber\\&=-(\tensor{ e}{^{A_0}_\mu}d\eta)^2+\delta_{{A_i}{A_j}}\tensor{e}{^{A_i}_{\mu}}\tensor{e}{^{A_j}_{\nu}}dx^\mu dx^\nu\,.
\end{align}
where in the last two steps we introduced two different choices of vierbeins. 

The first choice of vierbein corresponds to simply shifting the time $\eta$ but not boosting the tetrad, such that
\begin{align}
 \tensor{\tilde e}{^{A_0}_\mu}\approx (1+\delta\eta')\delta_\mu^0+\delta_\mu^i\partial_i \delta \eta\quad{\rm and}\quad \tensor{\tilde e}{^{A_i}_{\mu}}\approx \tensor{\delta}{^{A_i}_i}\,.
 \end{align} 
 However, the time-like vierbein is not identified with the direction of proper time of the particle and, therefore, the time-component of the local momenta associated with this vierbein is not the observed energy of the particle. In order to correct for this and use the time-like tetrad associated with the proper time, as we did in our general formulation (see Sec.~\ref{sec:Kinetictheory}), we have to introduce the second set of vierbeins in Eq.~\eqref{eq:dslocaltransf}, which are given by
\begin{align}
\tensor{e}{^{A_0}_\mu}\approx (1+\delta\eta')\delta_\mu^0\quad{\rm and}\quad \tensor{e}{^{A_i}_{\mu}}\approx \partial_i\delta \eta \delta_\mu^0\tensor{\delta}{^{A_i}^i}+\tensor{\delta}{^{A_i}_i}\,.
\end{align}
Thus, in the Kinetic Theory approach to a gas of particles, the time translation has Lorentz boosted the local vierbeins, namely
\begin{align}
 \tensor{ \tilde e}{^{A_0}_\mu}=\tensor{ e}{^{A_0}_\mu}+\partial_i \delta \eta\delta^{ij}\tensor{ e}{^{A_i}_j}=\tensor{\Lambda}{^{A_0}_{B}}\tensor{ e}{^{B}_\mu}\,,
\end{align} 
where
\begin{align}
\tensor{\Lambda}{^{A_0}_{A_0}}=1\quad{\rm and}\quad \tensor{\Lambda}{^{A_0}_{A_i}}=\partial_i\delta \eta \tensor{\delta}{_{A_i}^i}\,.
\end{align}

From the above argument we conclude that while $Q^{A}=a(\eta) P^{A}$ is time-independent at the background level, we have that at linear order in $\delta\eta$, $P^{A}$ rotates as $P^{A}\to P^{B}\tensor{\Lambda}{_B^{A}}$. Thus, we have that
\begin{align}
Q^{A_0}=a(\eta) P^{A_0}\to Q^{A_0}=a(\eta+\delta \eta) P^{B}\tensor{\Lambda}{_B^{A_0}}\approx Q^{A_0}(1+{\cal H}\delta \eta+\hat P^i\partial_i\delta\eta)\,.
\end{align}
This precisely recovers the gauge transformation of $Q^{A_0}$  \eqref{eq:tildeQA0} that we derived rigorously in Sec.~\ref{sec:common_uniform}, further supporting the separate universe approach. Thus, on super-Hubble scales (where we can neglect gradients terms), we simply have that
\begin{align}
Q^{A_0}\to Q^{A_0}(1+{\cal H}\delta \eta)\,,
\end{align}
and, therefore, it follows that
\begin{align}\label{eq:transformationfseparate}
f(Q^{A_0}) \to f(Q^{A_0}(1+{\cal H}\delta \eta))= f(Q^{A_0}) +{\cal H}\delta \eta Q^{A_0}\frac{\partial f}{\partial Q^{A_0}}\,.
\end{align}
Inserting the transformation rules of $f$ \eqref{eq:transformationfseparate} into the definition of $\Gamma$ \ref{eq:gammadef}, and using the solution for $\delta\eta$ in the Newton gauge \eqref{eq:deltaeta}, we arrive at
\begin{align}
\Gamma^{\rm N}=-{\cal H}\delta\eta^{\rm N}=-\frac{\Phi}{2}\,.
\end{align}
This result coincides with the strong adiabatic initial conditions given by \eqref{eq:stronggammaN}. Note that that Ref.~\cite{Dimastrogiovanni:2022eir} used the intuitive expectation that adiabatic initial conditions should be frequency independent to derive precisely such adiabatic initial conditions.

One may also wonder how the separate universe approach affects the integrated energy density, since it scales as $\rho\propto a^{-4}$. However, note that $\rho$ given by Eq.~\eqref{eq:rhoanddeltarho} explicitly depends on $P^{A_0}$ and $f(Q^{A_0})$, that is
\begin{align}
\rho=a^{-4}\int d\Omega \,dQ^{A_0} \,(Q^{A_0})^3\,f(Q^{A_0})=\int d\Omega \,dP^{A_0} \,(P^{A_0})^3\,f(Q^{A_0})\,,
\end{align}
where in the second step we used that $Q^{A_0}=aP^{A_0}$. In this form, we see that the separate universe approach also yields that $\delta\rho$ is related only to $\delta f$ as in Eq.~\eqref{eq:rhoanddeltarho}, consistent with our formulation.

Lastly, note that the weak adiabatic initial condition is also compatible with the separate universe approach if internal isocurvature fluctuations are already present before the time translation. Namely, the separate universe approach allows for an internal isocurvature fluctuation in the ``background'' homogeneous universe, that is
\begin{align}
\rho&= a^{-4}\int d\Omega \,dQ^{A_0} \,(Q^{A_0})^3\,f(Q^{A_0})+a^{-4}\int d\Omega \,dQ^{A_0} \,(Q^{A_0})^3\,\delta f_{\rm iso}(Q^{A_0})\nonumber\\&=a^{-4}\int d\Omega \,dQ^{A_0} \,(Q^{A_0})^3\,f(Q^{A_0})\,,
\end{align}
where in the last step we used Eq.~\eqref{eq:weakgammaiso}. Since the internal isocurvature $\delta f_{\rm iso}$ is already a first order quantity, it will not be affected by the time translation at first order in perturbation theory. With this, we conclude that both strong and weak adiabatic initial conditions given respectively by Eqs.~\eqref{eq:strongud} and \eqref{eq:weakud} (see also Eqs.~\eqref{eq:stronggammaN} and \eqref{eq:weakgammaN}) obtained from requiring a common uniform density slice are consistent with the separate universe approach.

\section{Conclusions \label{sec:conclusions}}

In cosmology, one generally expects that fluctuations generated during single field inflation start, at the onset of the Hot Big Bang, with adiabatic initial conditions on super-Hubble scales. Adiabatic initial conditions are defined by the absence of isocurvature fluctuations. In this work, we presented a complete classification of the possible adiabatic initial conditions for a gas of relativistic particles.

We showed that there are two main types of adiabatic initial conditions: the strong adiabatic initial condition, where fluctuations of the distribution function vanish in uniform density slices, and the weak adiabatic initial conditions, where there are internal isocurvature fluctuations that only vanish upon momentum integration. These two possibilities are explicitly given by Eqs.~\eqref{eq:strongud} and \eqref{eq:weakud} in the uniform density slice and by  Eqs.~\eqref{eq:stronggammaN} and \eqref{eq:weakgammaN} in the Newton gauge. We also showed in Sec.~\ref{sec:separate_universe}, that the same adiabatic initial conditions can be derived from the separate universe approach. Note that in recent studies of cosmic GW background anisotropies both possibilities have been considered, respectively, in Refs.~\cite{Dimastrogiovanni:2022eir,Malhotra:2022ply} and Refs.~\cite{Ricciardone:2021kel,Schulze:2023ich}. One must, therefore, study the GW generation mechanism in detail before setting a fixed adiabatic initial condition.

We have also highlighted the importance of describing the energy density of the particle gas as an integral over the local momentum of the particles \cite{Ma:1995ey} (see Eq.~\eqref{eq:rhoanddeltarho}), which is defined in the tangent space using the local tetrads. Furthermore, the phase space distribution function is also more appropriately described in terms of the local momentum, resulting in a simplified Boltzmann equation \eqref{eq:gammaequation}. If one used the spacetime momentum components instead, the density fluctuations of the particle gas are not directly related to fluctuations of the distribution function but include metric fluctuations as well (see Eq.~\eqref{eq:rhononlocal}). Although the latter contribution in the energy density fluctuations seem to violate standard adiabatic initial conditions \cite{ValbusaDallArmi:2023nqn,ValbusaDallArmi:2024hwm,Mierna:2024pkh}, we showed that it disappears if one uses the local momentum. We thus conclude that a gas of gravitons can also have adiabatic initial conditions.

In this work, we have restricted ourselves to collisionless relativistic particles. It would be interesting to extend the discussion to general types of particles and to include interactions as in Ref.~\cite{Malik:2004tf}. It would also be interesting to study concrete generation mechanisms of relativistic particles with adiabatic initial conditions and investigate which models are capable of introducing internal isocurvature, in particular for cosmic GWs. Lastly, the definition of the energy momentum tensor for low frequency GWs is subject to gauge ambiguities at higher order in cosmological perturbation theory. This could impact the identification of a graviton distribution function in the low momentum regime. We leave a more detailed study for future work.

\begin{acknowledgments}
I would like to thank M.~Bartelmann, N.~Bartolo, A.~Ganz, D.~Giulini, K.~Inomata, S.~Matarrese, A.~Ricciardone and M.~Sasaki for their helpful discussions. I specially thank J.~Lesgourgues for proposing to study the separate universe approach from the kinetic theory perspective. I also thank E.~Dimastrogiovanni, M.~Fasiello, A.~Malhotra and G.~Tasinato for participating in early stages of this work. This research is supported by the DFG under the Emmy-Noether program, project number 496592360, and by the JSPS KAKENHI grant No. JP24K00624.
\end{acknowledgments}

\appendix 

\section{Canonical momentum and geodesics with spacetime indices\label{app:appendix}}

Here we present for completeness some useful formulas in spacetime components. First, the Lagrangian for a point particle can be taken to be
\begin{align}
L=\frac{1}{2}g_{\mu\nu}\frac{dx^\mu}{d\lambda}\frac{dx^\nu}{d\lambda}\,,
\end{align}
where for a massless particle we must further impose $g^{\mu\nu}p_\mu p_\nu=0$ at the level of the equations of motion. The action can then be written as $S=\int d\lambda \,L(x^\mu,{dx^\nu}/{d\lambda})$. From this, it follows that the canonical conjugate momentum is given by
\begin{align}
p_\mu=g_{\mu\nu}\frac{dx^\nu}{d\lambda}\,.
\end{align}
The Hamilton equations then yield the geodesics equation, namely
\begin{align}
\frac{dp_\mu}{d\lambda}=g^{\alpha\beta}\Gamma^\sigma_{\mu\alpha}p_{\beta}p_{\sigma}\,.
\end{align}
And the Boltzmann equation for the distribution function of a gas of particles reads
\begin{align}\label{eq:Boltzmannspace-time}
\left[p^\mu \frac{\partial }{\partial x^\mu}+\frac{dp_\mu}{d\lambda}\frac{\partial }{\partial p_\mu}\right]f=\left[p^\mu \frac{\partial }{\partial x^\mu}+\Gamma^\sigma_{\mu\alpha}p^{\alpha}p_{\sigma}\frac{\partial }{\partial p_\mu}\right]f=0\,.
\end{align}

In cosmology, it is convenient to do a conformal transformation of the metric given by
\begin{align}
g_{\mu\nu}=\Omega^2(x^\alpha)\breve g_{\mu\nu}\,.
\end{align}
In that case, the Christoffel symbols change according to $\Gamma^\sigma_{\alpha\beta}=\breve\Gamma^\sigma_{\alpha\beta}+C^\sigma_{\alpha\beta}$ where $C^\sigma_{\alpha\beta}=\delta^\sigma_\alpha\partial_\beta\ln\Omega+\delta^\sigma_\beta\partial_\alpha\ln\Omega-\breve g_{\alpha\beta}\breve g^{\sigma\lambda}\partial_\lambda\ln\Omega$.
A quick calculation then shows that the geodesic equation becomes
\begin{align}
\frac{d\breve p_\mu}{d\Lambda}=\breve g^{\alpha\beta}\breve \Gamma^\sigma_{\mu\alpha}\breve p_{\beta}\breve p_{\sigma}\,,
\end{align}
where $d\Lambda=\Omega^2d\lambda$ and $\breve p_\mu=p_\mu$. Note that despite the equality, the index of $\breve p_\mu$ is raised with $\breve g^{\mu\nu}$, while that of $p_\mu$ with $g^{\mu\nu}$. Namely,
\begin{align}
\breve p_\mu=\breve  g_{\mu\nu}\frac{dx^\nu}{d\Lambda}\,,
\end{align}
is the canonical momentum of $x^\mu$ in the $\breve g_{\mu\nu}$ spacetime. Thus, Eq.~\eqref{eq:Boltzmannspace-time} can also be written in terms of tilded quantity and takes exactly the same form, explicitly
\begin{align}
\left[\breve p^\mu \frac{\partial }{\partial \breve  x^\mu}+\breve \Gamma^\sigma_{\mu\alpha}\breve p^{\alpha}\breve p_{\sigma}\frac{\partial }{\partial\breve  p_\mu}\right]f=0\,.
\end{align}
This latter equation is convenient as there is no scale factor in the new Christoffel symbols.

\bibliography{ref.bib}

\end{document}